\documentclass[%
reprint,
superscriptaddress,
 amsmath,amssymb,
 aps,
]{revtex4-2}

\usepackage{graphicx}
\usepackage{dcolumn}
\usepackage{bm}
\usepackage{hyperref}
\usepackage{xcolor}
\usepackage[caption=false]{subfig}
\usepackage[T1]{fontenc}

\def\1{\mathbf{1}}

\def\NN{{\mathbb N}}

\def \0{{\mathbf{0}}}

\begin{document}



\title{The threshold model with anticonformity under random sequential updating}
\author{Bartłomiej Nowak}
\email{bartlomiej.nowak@pwr.edu.pl}
\affiliation{%
Department of Theoretical Physics, Faculty of Fundamental Problems of Technology, Wrocław University of Science and Technology, 50-370 Wrocław, Poland
}%
\author{Michel Grabisch}
\email{michel.grabisch@univ-paris1.fr}
\affiliation{
University of Paris I Panthéon-Sorbonne, Paris School of Economics
106-112, Bd de l'Hôpital, Paris, France
}%
\author{Katarzyna Sznajd-Weron}
\email{katarzyna.weron@pwr.edu.pl}
\affiliation{%
Department of Theoretical Physics, Faculty of Fundamental Problems of Technology, Wrocław University of Science and Technology, 50-370 Wrocław, Poland
}%


\date{\today}

\begin{abstract}
We study an asymmetric version of the threshold model with anticonformity under asynchronous update mode that mimics continuous time.  We study this model on a complete graph using three different approaches: mean-field approximation, Monte Carlo simulation, and the Markov chain approach. The latter approach yields analytical results for arbitrarily small systems, in contrast to the mean-field approach, which is strictly correct only for an infinite system. We show that for sufficiently large systems, all three approaches produce the same results, as expected. We consider two cases: (1) homogeneous, in which all agents have the same tolerance threshold, and (2) heterogeneous, in which the thresholds are given by a beta distribution parametrized by two positive shape parameters $\alpha$ and $\beta$. The heterogeneous case can be treated as a generalized model that reduces to a homogeneous model in special cases. We show that particularly interesting behaviors, including social hysteresis and critical mass, arise only for values of $\alpha$ and $\beta$ that yield the shape of the distribution observed in real social systems.
\end{abstract}

\maketitle


\section{Introduction}
Within the broad class of two-state dynamics \cite{Gle:13}, threshold models are particularly useful for describing various social and economical phenomena \cite{Gra:78,Wat:02,Bre:17}. As other binary-state opinion dynamics \cite{Jed:Szn:19}, the threshold model describes the social influence in decision-making for the choice between precisely two alternatives, often denoted by $1$ (agree, adopt, be active, etc.) and $0$ (disagree, refuse, be inactive, etc.). Although a binary decision framework seems to be oversimplified, it is relevant to surprisingly many complex problems \cite{Wat:02}. 

In the original threshold models of collective behavior, proposed by Schelling \cite{Sch:78} and Granovetter \cite{Gra:78}, an agent takes action $1$  if the proportion of his neighbors in state $1$ exceeds some threshold, otherwise action $0$ is taken. It means that an agent at state $1$ may return to state $0$, because not enough neighbors are active. On the other hand, in many other threshold models, the transition from state $1$ to state $0$ is forbidden \cite{Wat:02,Dod:Wat:04,Juu:Por:19}. 

Here we will use the original formulation, in which a transition from $1$ to $0$ is possible, as in \cite{Gra:78,Lee:Hol:17,Gra:Li:20}, but additionally in the presence of anticonformity. Such a model has been already studied from a mathematical point of view under the synchronous update mode \cite{Gra:Li:20}. The study was focused on finding absorbing classes, cycles, etc. In this paper, we investigate the same model but under random sequential updating, which mimics continuous time. Contrarily to \cite{Gra:Li:20},
we focus on phase transitions and phase diagrams, which is a typical approach for statistical physics of opinion formation \cite{Cas:For:Lor:09,Gal:Mar:15,Rad:etal:18,Cal:Cro:Pen:19,Vie:etal:20}. 

We study the model on the complete graph, which enables to obtain exact  results within the mean-field approach. Independently, we conduct Monte Carlo simulations to validate the theoretical approach. Finally, we present a Markov Chain approach, which not only allows us to obtain results for arbitrary small systems, but also to derive the stationary distribution of visited states. 

\section{Model}
We consider a society of $n$ agents placed at the vertices of an arbitrary graph $G = (N, E)$, where $N = \{1, . . . , n\}$ is a set of vertices (agents) and $E$ is the set of undirected edges. Each agent $i$ has a set of neighbors $K_i = \{j \in N : \{i , j\} \in E\}$, and cardinality of this set $|K_i|=k_i$ is the degree of agent $i$. As in many other models, an agent can be in one of two alternative states: $1$ (agree, adopt,
be active, etc.) or $0$ (disagree, refuse, be inactive, etc.). Following \cite{Gra:Li:20}, we use the term “active” for agents in state $1$, and “inactive” for agents in state $0$, and denote by $a_i(t)$ the state (action) taken by agent $i$ at time $t$.

We consider two types of social response, anticonformity and conformity, occurring with complementary probabilities $p$ and $1-p$ respectively. In both cases, an agent can change its state, if the ratio of active neighbors is above its tolerance threshold $r_i \in [0,1]$. Threshold $r_i$ of each agent is the realization of the random variable $R$ with arbitrary distribution function $F_{R}(r)$ and does not change in time. 
In case of conformity, an agent follows the others, whereas in case of anticonformity he takes an opposite state to others. Therefore, the dynamics of the agent's state in case of conformity can be written as \cite{Gra:Li:20}:
\begin{equation}
    a_{i}(t+\Delta t) = \begin{cases}
    1, & \text{if } \frac{1}{k_i} \sum\limits_{j \in K_{i}} a_{j}(t) \geqslant r_i \\
    0, & \text{otherwise},
  \end{cases} \\
\label{eq:a_conf}
\end{equation}
whereas in case of anticonformity \cite{Gra:Li:20}:
\begin{equation}
    a_{i}(t+\Delta t) = \begin{cases}
    0, & \text{if } \frac{1}{k_i} \sum\limits_{j \in K_{i}} a_{j}(t) \geqslant r_i \\
    1, & \text{otherwise}
  \end{cases} \\
\label{eq:a_anti}
\end{equation}

In this paper, we use the random sequential update mode, which means that an elementary update consists of:
\begin{enumerate}
    \item random drawing of agent $i$ from all $n$ agents
    \item with probability $p$ agent $i$ anticonforms to the neighborhood, i.e.,  takes action $a_{i}(t+\Delta t)$ according to Eq. (\ref{eq:a_anti})
    \item with complementary probability $1-p$ agent $i$ conforms to the neighborhood, i.e., takes action $a_{i}(t+\Delta t)$ according to Eq. (\ref{eq:a_conf})
    \item time is updated: $t:=t+ \Delta t$
\end{enumerate}
As usually, $\Delta t=1/n$ which means that the time unit consists of $n$ elementary updates, which corresponds to one Monte Carlo step (MCS).

\section{Transition probabilities}
Since we limit our study to the complete graph, we can fully describe the state of the system by a single random variable:
\begin{equation}
    c = \frac{n_{1}}{n},
\end{equation}
where $n_{1}$ is the number of agents in state $1$ and thus $c$ is the ratio of active agents. Therefore, there are $n+1$ possible states of the system: $0,\frac{1}{n},\frac{2}{n},\ldots,1$.

Because we use the sequential (asynchronous) update mode, at most one agent can change its state at a time, and thus we can introduce the following transition probabilities:
\begin{align}
  \gamma^{+}(c) &= Pr\left(c(t + \Delta t) = c(t) + \frac{1}{n}\right), \nonumber \\
  \gamma^{-}(c) &= Pr\left(c(t + \Delta t) = c(t) - \frac{1}{n}\right).
  \label{eq:transition_rates}
\end{align}

For our model, the explicit form of these probabilities can be written, according to the algorithm described in the previous section, as follows:
\begin{align}
  \gamma^{+}(c) &= (1-p)(1-c)Pr(R \leqslant c) + p(1-c)Pr(R > c), \nonumber \\
  \gamma^{-}(c) &= (1-p)cPr(R > c) + pcPr(R \leqslant c),
  \label{eq:gammasgeneral}
\end{align}
where $Pr(R \leqslant c)$ is the probability that the concentration $c$ of active agents is bigger than or equal to the threshold $R$ of the considered agent. This probability is simply the value of the cumulative distribution function $F_{R}(r)$ at $r=c$. Similarly, $Pr(R > c)$ is the probability that the concentration of active voters does not exceed the threshold of considered agents and thus it is equal to $1 - F_{R}(c)$. Therefore, we obtain
\begin{align}
  \gamma^{+}(c) &= (1-p)(1-c)F_{R}(c) + p(1-c)(1-F_{R}(c)), \nonumber \\
  \gamma^{-}(c) &= (1-p)c(1-F_{R}(c)) + pcF_{R}(c).
  \label{eq:general}
\end{align}

As can be seen from Eq.(\ref{eq:transition_rates}), the concentration of active agents $c$ is a random variable. However, we can easily write the evolution equation for the expected value of $c$. Moreover, for $n \rightarrow \infty$ we can assume that $c$ localizes to the expectation value. Therefore, we can write \cite{Jed:Szn:19}:
\begin{equation}
    \frac{dc}{dt} =  \gamma^{+}(c) - \gamma^{-}(c) 
    \label{eq:dcdt}
\end{equation}
As usually, we focus on the steady states, i.e., those for which 
\begin{equation}
    \frac{dc}{dt} =  0.
    \label{eq:stationary}
\end{equation}
In the next two sections, we will use the condition (\ref{eq:stationary}) to calculate the stationary concentration of active agents for two cases: (1) the homogeneous one, in which all agents have the same tolerance threshold (2) the heterogeneous one, in which the distribution of thresholds $F_{R}(r)$ is given by the beta distribution.
We will compare the analytical results with the results of Monte Carlo simulations for the system of size $n=10^4$, averaged over $10$ independent runs collected after $10^4$ Monte Carlo steps. For the Monte Carlo simulations, two types of initial conditions will be used to reproduce all stable solutions of Eq. (\ref{eq:stationary}): (1) all agents initially active, which will be denoted by $c(0)=1$ and (2) all agents initially inactive, which will be denoted by $c(0)=0$.

\section{One threshold}
\begin{figure}[htp]
    \includegraphics[width=\linewidth]{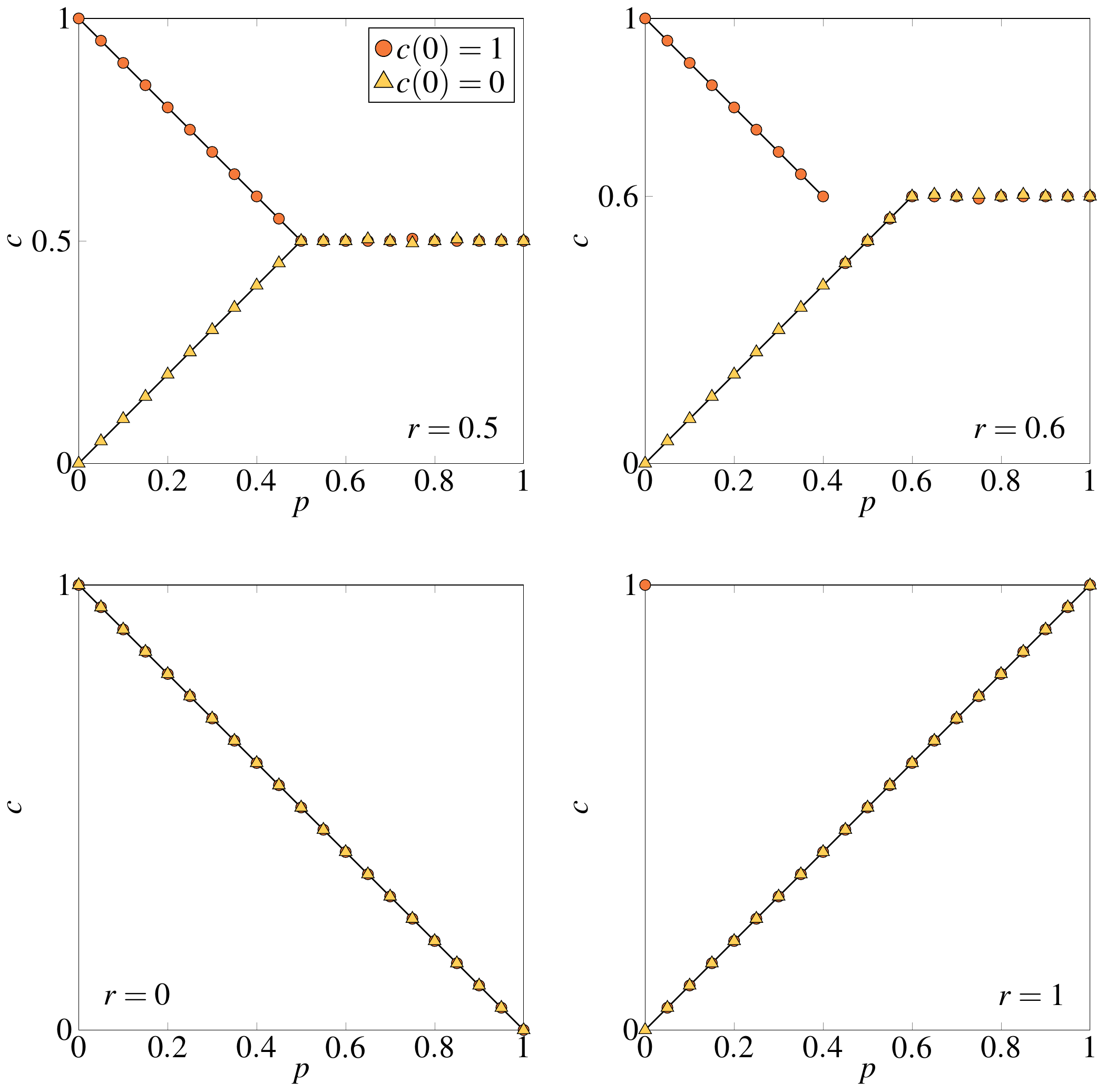}
    \caption{Dependency between the stationary concentration of active agents $c$ and the probability of anticonformity $p$ for model with one threshold for different values of the parameter $r$ (indicated in the plots). Solid lines represent stable fixed points obtained analytically from Eq.~(\ref{eq:cponethreshold}). Symbols represent Monte Carlo simulations from two initial conditions indicated in the legend.}
    \label{fig:fig1}
\end{figure}

\begin{figure*}[htp]
    \includegraphics[width=\linewidth]{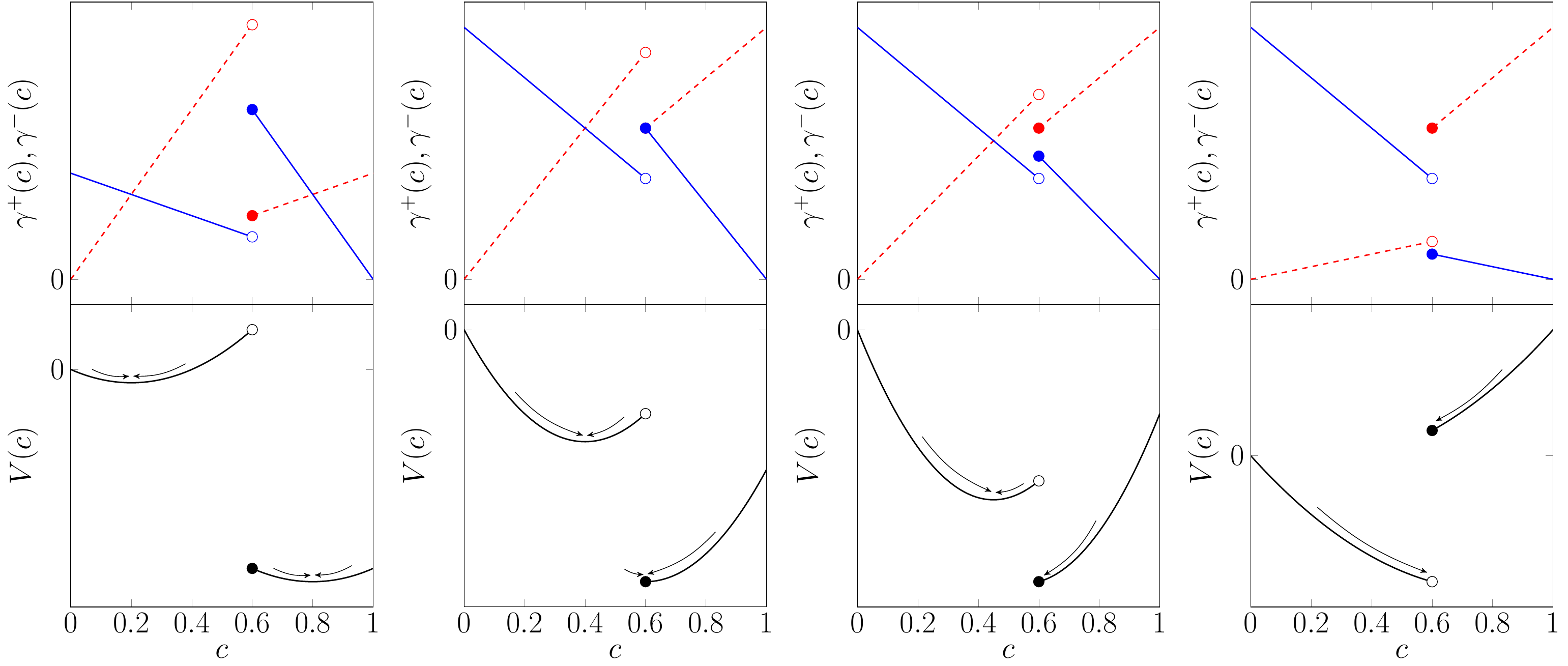}
    \includegraphics[width=\linewidth]{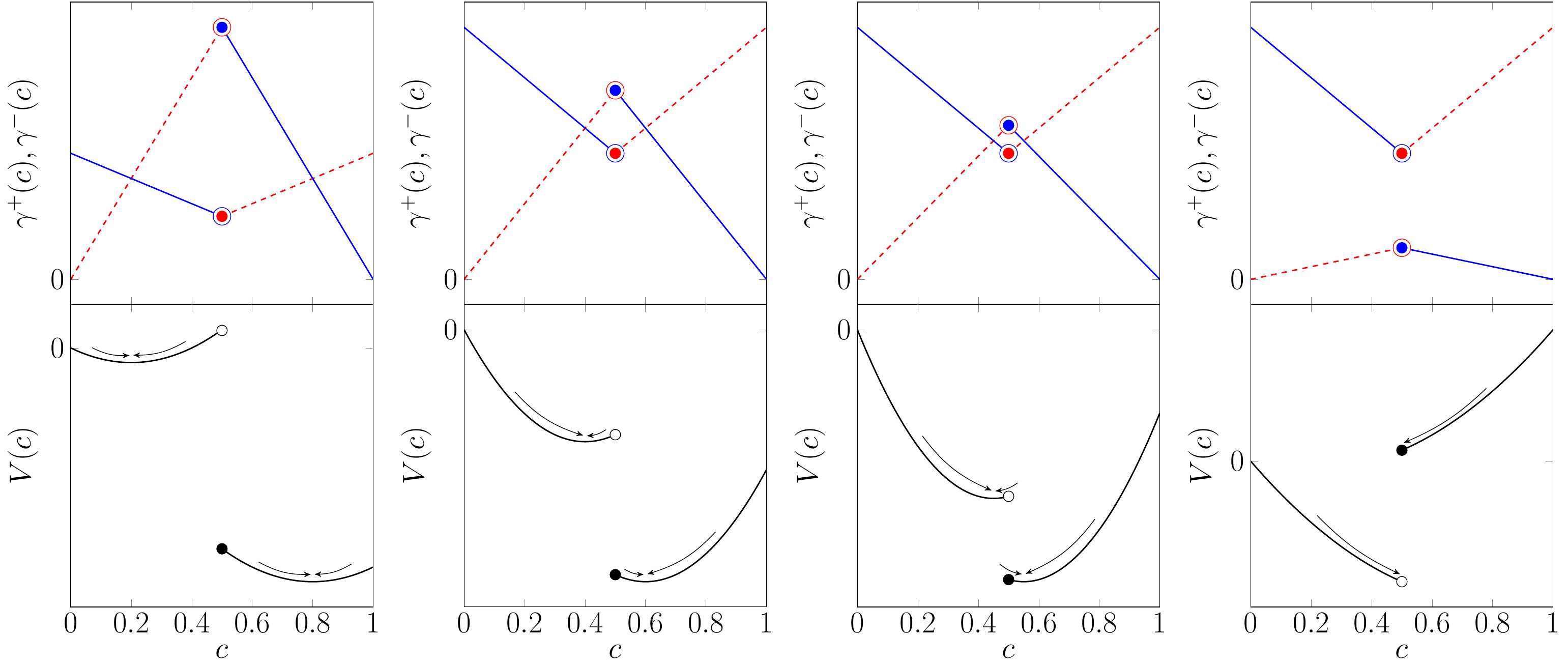}
    \caption{Analysis of the steady states and the stability of the system for two values of threshold $r=0.6$ (two first rows) and $r=0.5$ (two last rows) and four values of $p = 0.2$ (first column), $p = 0.4$ (second column), $p = 0.45$ (third column), $p = 0.8$ (fourth column). In first and third row, solid lines represents values of $\gamma^{+}$ and dotted lines stands for $\gamma^{-}$ obtained with Eqs. (\ref{eq:gammassplit}). Potentials $V(c)$ (second and fourth row) are obtained with Eqs. (\ref{eq:potentials}). In all subplots, filled circles denote continuity, while empty circles denote lack of continuity at this point.}
    \label{fig:potgamm}
\end{figure*}

In this case, the random variable $R$ takes one value for all agents in the system, i.e., all voters have the same threshold $r$:
\begin{align}
  F_{R}(c) &= \mathbf{1}_{ \{ r \leqslant c\}} , \nonumber \\ 
  1-F_{R}(c) &= \mathbf{1}_{ \{ r > c\}},
\label{eq:FR_one}
\end{align}
where $\mathbf{1}_{ \{ r \leqslant c\}} = 1$ when $r \leqslant c$ and 0 otherwise. Inserting (\ref{eq:FR_one}) to Eq.(\ref{eq:general}) and then to Eq. (\ref{eq:dcdt}) we obtain:
\begin{align}
    \frac{dc}{dt} = (1-p)\left[(1-c)\mathbf{1}_{ \{ r \leqslant c\}} - c\mathbf{1}_{ \{ r > c\}}\right] \nonumber \\
    + p\left[(1-c)\mathbf{1}_{ \{ r > c\}} - c\mathbf{1}_{ \{ r \leqslant c\}}\right]
    \label{eq:generalonethreshold}
\end{align}
From (\ref{eq:generalonethreshold}), we obtain several trivial fixed points:
\begin{align}
    &(p = 0, \; c = 0)  && \forall {r \neq 0} \\
    &(p = 1, \; c = 0) && r = 0 \\
        &(p = 0, \; c = 1)  && \forall {r \in [0,1]} \\
    &(p = 0.5, \; c = 0.5)  && \forall {r \in [0,1]} \\
    &(p = 1-r, \; c = r)  && \forall {r \in [0,1]} \label{eq:fixedpoint5}
\end{align}

The remaining solutions can be obtained by solving Eq. (\ref{eq:stationary}), which leads to
\begin{equation}
    p = \frac{\mathbf{1}_{ \{ r \leqslant c\}} - c}{\mathbf{1}_{ \{ r \leqslant c\}} - \mathbf{1}_{ \{ r > c\}}},
    \label{eq:cponethreshold}
\end{equation}
what is equivalent to the following cases
\begin{equation}
    \forall{c < r}\ \ p = c, \quad \forall{c\geqslant r}\ \ p = 1-c.
\end{equation}
From the above analysis we do not obtain steady state for any value of ${r\in[0,1]}$ if $p > r$. However, from the evolution of Eq. (\ref{eq:generalonethreshold}), as well as from the Monte Carlo simulations, it seems that the system approaches the state $c = r$ for $p > r$, which is shown in Fig.~\ref{fig:fig1}. This raises the question of what the evolution of the system actually looks like and how it can be understood on the basis of transition probabilities $\gamma^+(c),\gamma^-(c)$. The above question can be answered if we split the transition probabilities into cases
\begin{align}
    &\forall{c<r}\ \ \gamma^+(c)=p(1-c)\ \wedge\ \gamma^-(c)=(1-p)c, \nonumber \\
    &\forall{c\geqslant r}\ \ \gamma^+(c)=(1-p)(1-c)\ \wedge\ \gamma^-(c)=pc.
    \label{eq:gammassplit}
\end{align}
Doing this we easily observe that they do not cross at any point, when $\forall{r>0.5}\ \ p\geqslant r$ or $\forall{r\leqslant 0.5}\ \ p > r$, see the fourth column of Fig.~\ref{fig:potgamm}. This implies no steady state. On the other hand, for $p<r$ transition probabilities $\gamma^-(c)$ and $\gamma^+(c)$ cross each other, as shown in the first three columns of Fig.~\ref{fig:potgamm}, i.e., the steady state $\gamma^-(c)=\gamma^+(c)$ exists.

There is another illustrative way to visualize the dynamics of the system based on the idea of potential $V(c)$ \cite{Str:15}:
\begin{equation}
   V(c) = - \int F(c) dc = - \int \frac{dc}{dt} dc,
\end{equation}
where $F(c)$ plays the role of a generalized force, which drives the dynamics of the system. Using such an approach, we draw a ball sliding down the walls of a potential well \cite{Str:15}, as shown in Fig.~\ref{fig:potgamm}. To calculate the explicit form of $V(c)$ we use Eq.~(\ref{eq:gammassplit}), which leads to:
\begin{align}
    &\forall{c<r}\ \ V(c) = \frac{c^{2}}{2} - cp, \nonumber \\
    &\forall{c\geqslant r}\ \ V(c) = \frac{c^{2}}{2} - c(1 - p).
    \label{eq:potentials}
\end{align}
The steady states are local extrema of $V(c)$. From Eq. (\ref{eq:potentials}) we see that the potential has a discontinuity at $c = r$ which implies no maximum (unstable steady state). Still, at most two minima (stable steady states) are possible. In general, the number of minima, denoted by $M(r,p)$, can be described as follows
\begin{align}
    \forall{r>0.5}\ \ M(r,p) &=  
    \begin{cases}
    2, & \text{for } p \leqslant 1-r \\
    1, & \text{for } 1-r < p < r \\
    0, & \text{for } p \geqslant r,
  \end{cases} \\
  \forall{r\leqslant 0.5}\ \ M(r,p) &=  
    \begin{cases}
    2, & \text{for } p < r \\
    1, & \text{for } r \leqslant p \leqslant 1-r \\
    0, & \text{for } p > r.
  \end{cases}
\end{align}
In conclusion, despite the lack of steady state in the case $M(r,p) = 0$ we can observe the flow of the system is towards the point $c = r$. It reaches an asymptotic minimum at this point because from both, the left and right boundaries, the system flow is towards this minimum. This explains the behavior shown in Fig. \ref{fig:fig1}, which was at first incomprehensible and inspired the above analysis.

\section{Beta distribution}
\begin{figure}[htp]
    \includegraphics[width=\linewidth]{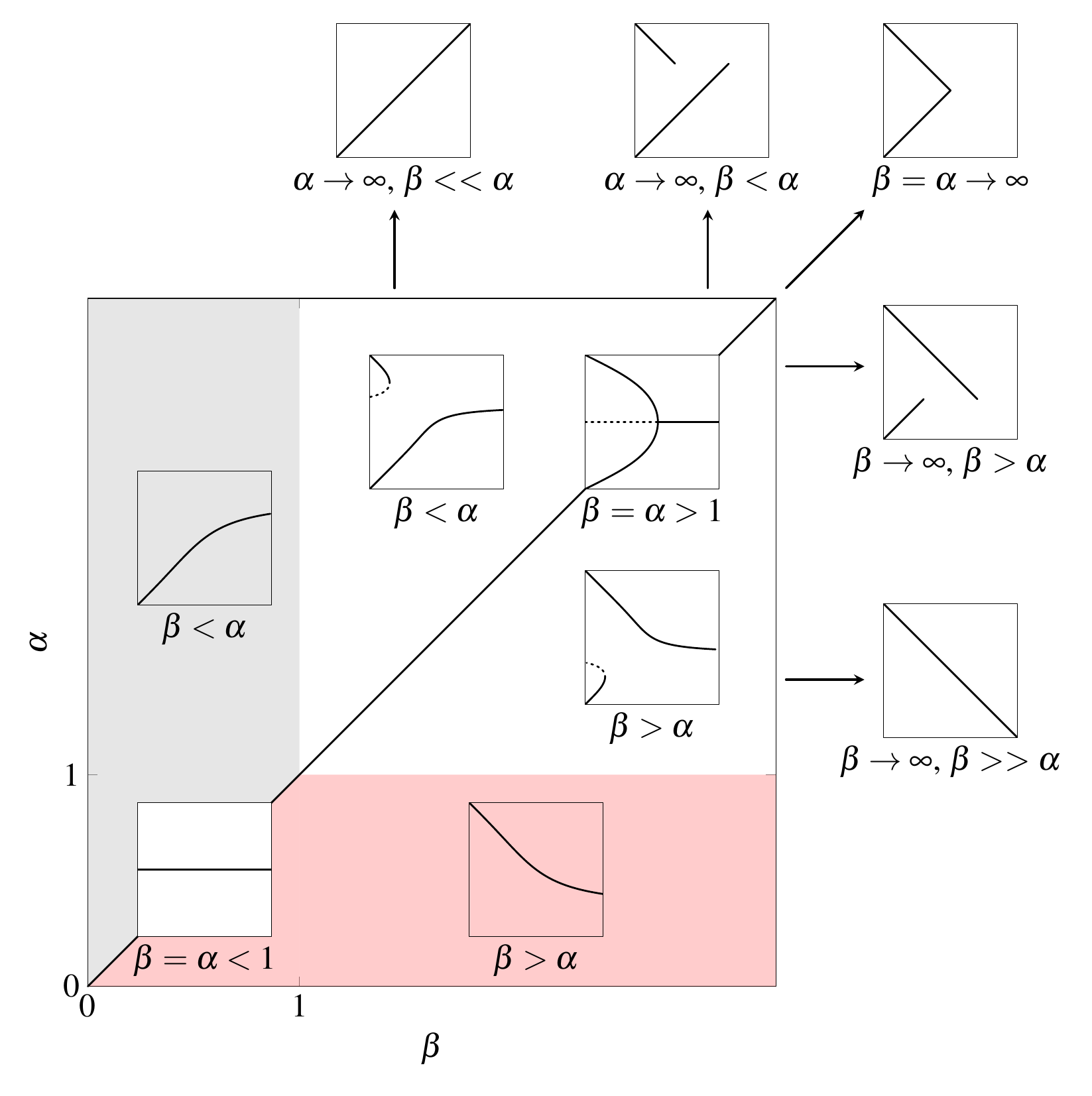}
    \caption{Phase diagram for the heterogeneous model with thresholds described by the beta distribution parametrized by two shape parameters $\alpha$ and $\beta$. Each inset shows representative behavior of $c(p)$ for a given area of the phase diagram. Solid lines in the insets correspond to stable stationary states, whereas dashed lines correspond to unstable stationary states.}
    \label{fig:phase}
\end{figure}

In the previous section, we studied the homogeneous system, in which all agents had the same value of the tolerance threshold $r$. However, we can also consider more general distributions of thresholds, allowing for heterogeneity. The most useful are distributions whose support values $r \in [0,1]$ and show a variety of shapes.
This is the case of the beta distribution with two parameters $\alpha$ and $\beta$, considered previously, for the models of tolerance without anticonformity \cite{Bre:17}. It has a well-defined cumulative distribution function
\begin{equation}
    F_{R}(r) = I_{r}(\alpha,\beta) = \frac{B(r,\alpha,\beta)}{B(\alpha,\beta)},
   \label{eq:betaprobabilities}
\end{equation}
where $I_{r}(\alpha,\beta)$ is the regularized incomplete beta function, which can be defined in terms of the incomplete beta function $B(r, \alpha,\beta)$ and the complete beta function $B(\alpha,\beta)$.
Inserting $F_{R}(r)$ given by Eq. (\ref{eq:betaprobabilities}) to (\ref{eq:general}) we obtain the transition probabilities $\gamma^+(c), \gamma^-(c)$. Then inserting them to Eq. (\ref{eq:dcdt}) we get
\begin{align}
    \frac{dc}{dt} = (1-p)\left[(1-c)I_{c}(\alpha,\beta) - c(1-I_{c}(\alpha,\beta))\right] \nonumber \\
    + p\left[(1-c)(1-I_{c}(\alpha,\beta)) - cI_{c}(\alpha,\beta)\right].
    \label{eq:betageneral}
\end{align}
Again, we can point out the obvious steady states $(p=0, \; c=0)$, $(p=0, \; c=1)$ for arbitrary values of $\alpha$ and $\beta$. For $c = 1/2$ formula (\ref{eq:betageneral}) boils down to
\begin{equation}
    \left.\frac{dc}{dt}\right|_{c = \frac{1}{2}} = \left(I_{\frac{1}{2}}(\alpha,\beta) - \frac{1}{2}\right)(1-2p),
\end{equation}
which has two roots. The first one $p=1/2$ gives the fixed point $(p = \frac{1}{2}, c = \frac{1}{2})$. The other root $I_{\frac{1}{2}}(\alpha,\beta) = \frac{1}{2}$ exists if the beta distribution is symmetric around the value $\frac{1}{2}$. This happens for $\alpha = \beta$, which leads to the conclusion that the value $c = \frac{1}{2}$ is a fixed point for all values of $p$ if $\alpha = \beta$. For all remaining solutions we have the following relation:
\begin{equation}
    p = \frac{I_{c}(\alpha,\beta) - c}{2I_{c}(\alpha,\beta) - 1}.
    \label{eq:betasolution}
\end{equation}
The information about the stability of the steady state is given by the sign of the derivative
\begin{equation}
    \frac{dF}{dc} = \frac{c^{\alpha - 1}(1-c)^{\beta - 1} \Gamma(\alpha + \beta)}{\Gamma(\alpha)\Gamma(\beta)}(1-2p) - 1.
\end{equation}
The state is stable if the above derivative is negative and unstable otherwise. The overall behavior of the model is summarized in Fig. \ref{fig:phase}. In the insets of this figure the dependence between the stationary value of $c$ and parameter $p$ is shown. Two shaded areas in Fig. \ref{fig:phase} correspond to the situation in which at least one of the parameters $\alpha, \beta$ is smaller than $1$. In this case, for $p>0$ there is always only one steady state and $c$ is monotonically increasing ($\beta < \alpha$), monotonically decreasing ($\beta > \alpha$) or constant ($\beta = \alpha$) function of $p$. 

Recalling the shape of the probability density function (PDF) of the beta distribution, we can draw some conclusions.
If the PDF of the tolerance threshold is a monotonically decreasing function of the threshold $r$, then the concentration of active agents decreases with the probability of anticonformity $p$, and vice versa. If the PDF has the highest values at $r=0$ and $r=1$, being a convex function of $r$, then for all values of $p>0$ the stationary value of active agents is $0.5$.

The most complex behavior is seen if both shape parameters $\alpha, \beta$ are greater than $1$ but not infinitely large, which corresponds to a uni-modal PDF, with zero probabilities at both end of the interval range, i.e., at $r=0$ and $r=1$. This case correspond to  moderate tolerance \cite{Bre:17}, and it is a typical shape of the distribution of actual trait manifestations in behavior, as reported by psychologists \cite{Fle:Gal:09}. In such a case, the phase transitions appear, as shown in Fig.~\ref{fig:phase}. As long as $\beta=\alpha$, which corresponds to the symmetric PDF, there is a continuous phase transition between the phase in which one type of agent (active or inactive) dominates, and the symmetrical phase without the domination.
The critical point, at which this transition occurs, can be calculated by solving the equation
    \begin{equation}
        \left.\frac{dF}{dc}\right|_{c = \frac{1}{2}} = 0,
    \end{equation}
which gives:
\begin{equation}
p_{1}^{*} = \frac{1}{2} - \frac{\Gamma^2(\alpha)}{2^{1-2(\alpha-1)}\Gamma(2\alpha)}.
\end{equation}

For $\alpha \neq \beta$, as long as shape parameters are finite and at least one of them is larger than 1, we obtain an interesting behavior, with the jump at some value of $p=p^*$ and hysteresis, as shown in Fig. \ref{fig:phase}. This can be especially useful to describe the innovation diffusion. For example, if  $\beta > \alpha$ then for the small value of $p<p^*$ there is possibility of high adoption if the initial fraction of adopted is above the critical mass. However, if the initial fraction of adopted is too low, i.e. below the critical mass, the innovation cannot spread in the society. Similar behavior has been recently reported for the completely different mathematical model of the collective decision-making with social learners for unequal merit options \cite{Yan:etal:21}.

It is worth noticing that for $\alpha,\beta \rightarrow \infty$ we can recover the solution for the model with one threshold, as shown in Fig. \ref{fig:phase}. We are able to do that by recalling the formula giving the mode of the beta distribution with $\alpha, \beta > 1$:
\begin{equation}
m = \frac{\alpha - 1}{\alpha + \beta - 2}.
\end{equation}
While $\alpha,\beta \rightarrow \infty$, the beta distribution is a 1-point degenerate distribution with probability 1 at the midpoint $m$ and 0 elsewhere. Thus, to obtain the case with the mode at the point $m = r$, i.e., recover the distribution for one threshold, parameters $\alpha$ and $\beta$ should follow the formula
\begin{equation}
\beta = \frac{(1-r)\alpha - 1 + 2r}{r}
\end{equation}
for $\alpha,\beta \rightarrow \infty$.

\begin{figure*}[htp]
    \subfloat{\includegraphics[width=\linewidth]{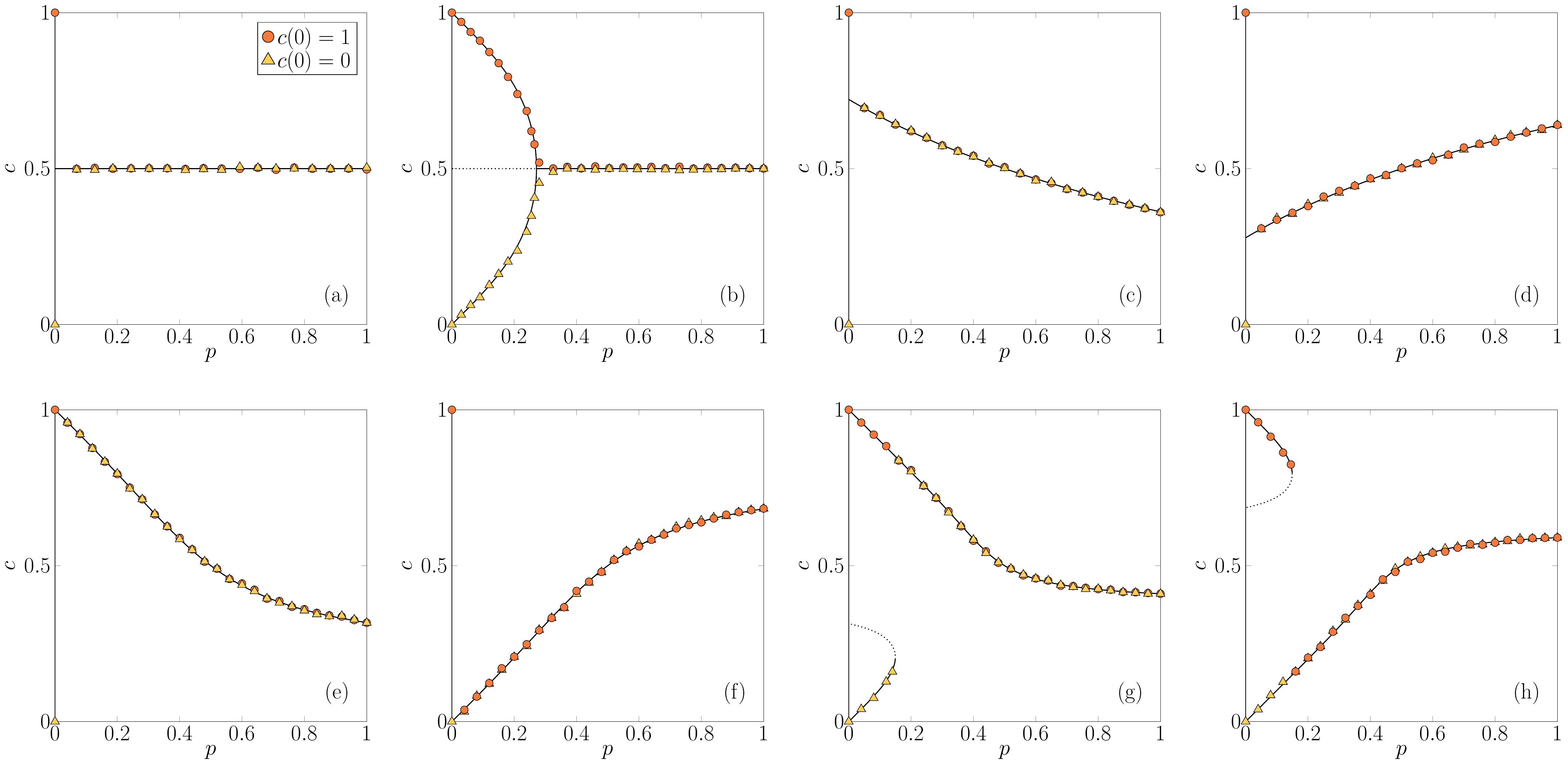}\label{fig:codp-a}}
    \subfloat{\label{fig:codp-b}}
	\subfloat{\label{fig:codp-c}}
	\subfloat{\label{fig:codp-d}}
	\subfloat{\label{fig:codp-e}}
	\subfloat{\label{fig:codp-f}}
	\subfloat{\label{fig:codp-g}}
	\subfloat{\label{fig:codp-h}}
    \caption{Representative dependencies between the stationary concentration of spins up and the probability of anticonformity for model with beta distribution for different values of the parameters $\alpha$ and $\beta$: (a) $\alpha = \beta \leqslant 1$, (b) $\alpha = \beta > 1$, (c) $\alpha < \beta < 1$; $\alpha$ close to $\beta$, (d) $1 > \alpha > \beta$; $\alpha$ close to $\beta$, (e) $\alpha \leqslant 1\ \wedge\ \alpha < \beta$, (f) $\beta \leqslant 1\ \wedge\ \alpha > \beta$, (g) $1 < \alpha < \beta$, (h) $\alpha > \beta > 1$. Solid and dotted lines represent stable and unstable steady states respectively, obtained with Eq.~(\ref{eq:betasolution}). The exact values of parameters in the plots are as
    follows: (a) $\alpha = \beta = 0.9$, (b) $\alpha = \beta = 4$, (c) $\alpha = 0.1$; $\beta = 0.2$, (d) $\alpha = 0.2$; $\beta = 0.1$, (e) $\alpha = 1$; $\beta = 3$, (f) $\alpha = 3$; $\beta = 1$, (g) $\alpha = 5$; $\beta = 8$, $(h)$ $\alpha = 8$; $\beta = 5$. Symbols represent Monte Carlo simulations from two initial conditions, denoted in the legend. The results are averaged over 10 runs and collected after $10^4$ MCS for system of size $10^4$.}
    \label{fig:codp}
\end{figure*}

All results obtained analytically for beta distribution can be also obtained by Monte Carlo simulations, as shown in Fig. \ref{fig:codp}.

\section{Markov Chain approach}
Previously, we were assuming that the size of the system is infinite, i.e., $n \rightarrow \infty$. However, such an assumption is not very realistic for social systems. Actually, social scientists are often interested in small systems. Therefore, in this section, we make analysis of the convergence of $c$ in the long run using Markov chains for arbitrary small systems. The advantage of the Markov chain approach in the context of agent-based modeling of opinion dynamics has been already reported in \cite{Ban:Li:Ara:12}. 

Transition probabilities given by Eq.~(\ref{eq:transition_rates}) allows us to write the $(n+1)\times(n+1)$ transition matrix, whose general term $(i,j)$ indicates the probability of transition from state $i$ to state $j$.
Due to the asynchronous update mode, $\mathbf{P}$ is a tridiagonal row-stochastic matrix:
\begin{widetext}
\begin{equation}
    \renewcommand{\arraystretch}{1.5}
    \mathbf{P} = \begin{bmatrix}
    \gamma^0(0) & \gamma^+(0) & 0 & 0 & \cdots & 0\\
    \gamma^-\left(\frac{1}{n}\right) & \gamma^0\left(\frac{1}{n}\right) & \gamma^+\left(\frac{1}{n}\right) & 0 & \cdots & 0\\
    0 & \gamma^-\left(\frac{2}{n}\right) & \gamma^0\left(\frac{2}{n}\right) & \gamma^+\left(\frac{2}{n}\right) & \cdots & 0\\
    0 & \cdots & \ddots & \ddots & \ddots & 0\\
    0 & \cdots & 0 & \gamma^-\left(\frac{n-1}{n}\right) & \gamma^0\left(\frac{n-1}{n}\right) & \gamma^+\left(\frac{n-1}{n}\right) \\
    0 & \cdots & 0 & 0 & \gamma^-(1) & \gamma^0(1)
    \end{bmatrix}
\end{equation}
\end{widetext}
with $\gamma^0(c)=\nobreak 1-\gamma^+(c)-\gamma^-(c)$. This process is a random walk process. Its transition graph is strongly connected and aperiodic, hence $\mathbf{P}$ is a primitive matrix, i.e., the only absorbing class is the set of all states. This means that in the long run, the system at time $t$ can be in any of the $n+1$ states, and there is no stabilization \cite{Kem:Sne:76,Sen:06}.

From Markov chain theory, the limit vector $\pi=\nobreak[\pi(0),\cdots,
\pi(c),\cdots,\pi(1)]$ giving the probability $\pi(c)$ to be in state $c$ in
the long run is obtained as the left eigenvector of $\mathbf{P}$ associated to
eigenvalue 1, i.e., $\pi$ is the solution of the system
\begin{align}
  (\mathbf{P}^T-I)z&=0 \nonumber \\
  1^Tz&=1
  \label{eq:Pmatrix}
\end{align}
From now on, to avoid heavy notation, we denote $\gamma^+(k/n)$ by
$\gamma^+(k)$, and similarly for $\gamma^-(k/n)$, $\pi(k/n)$, etc.
We obtain
\begin{widetext}
\begin{equation}
\renewcommand{\arraystretch}{1.5}
    \mathbf{P}^T-I = \begin{bmatrix}
  -\gamma^+(0) & \gamma^-(1) & 0 & 0 & 0 & \cdots & 0 \\
  \gamma^+(0) & -\gamma^-(1)-\gamma^+(1) & \gamma^-(2) & 0 & 0 & \cdots & 0 \\
  0 & \ddots & \ddots & \ddots & 0 & \cdots & 0\\
  0 & \cdots & \gamma^+(k-1) & -\gamma^-(k)-\gamma^+(k) & \gamma^-(k+1) & \cdots & 0\\
  0 & \cdots & 0 & \ddots & \ddots & \ddots & 0\\
  0 & \cdots & 0 & 0 &\gamma^+(n-2) & -\gamma^-(n-1)-\gamma^+(n-1) & \gamma^-(n)\\
  0 & \cdots & 0 & 0 & 0 & \gamma^+(n-1) & -\gamma^-(n)
  \end{bmatrix}
\end{equation}
\end{widetext}
Solving the system yields
\begin{align}
  \pi(0) &= \frac{\gamma^-(1)}{\gamma^+(0)}\pi(1)\nonumber\\
  \pi(1) & =
  \frac{\gamma^-(2)}{\gamma^+(1)}\pi(2)\nonumber\\
  \vdots & = \vdots \nonumber\\
  \pi(k)& = \frac{\gamma^-(k+1)}{\gamma^+(k)}\pi(k+1)\label{eq:k}\\
  \vdots &  = \vdots \nonumber \\
  \pi(n-1) & = \frac{\gamma^-(n)}{\gamma^+(n-1)}\pi(n).\nonumber
\end{align}
This yields:
\begin{equation}
    \pi(k) = \frac{\gamma^+(k-1)}{\gamma^-(k)}
    \frac{\gamma^+(k-2)}{\gamma^-(k-1)}\cdots \frac{\gamma^+(0)}{\gamma^-(1)}\pi(0) \quad (k=1,\ldots,n).
    \label{eq:pigeneral}
 \end{equation}
In the case of one threshold we are able to derive the above formulas analytically. Using Eqs. (\ref{eq:gammassplit}) we obtain:
\begin{align*}
\pi(k) & = \frac{(1-p)(k+1)}{p(n-k)}\pi(k+1) &   (k<rn-1)\\ 
\pi(k) & = \frac{k+1}{n-k}\pi(k+1) & (rn-1\leqslant k<rn)\\
\pi(k) & = \frac{p(k+1)}{(1-p)(n-k)}\pi(k+1) &   (k\geqslant rn).
\end{align*}
Let us find when $\pi(k)$ is increasing or decreasing.
Supposing $k<rn-1$, we have:
\begin{align*}
  \frac{(1-p)(k+1)}{p(n-k)}\leqslant 1 & \Leftrightarrow (1-p)(k+1)\leqslant p(n-k)\\
  & \Leftrightarrow k\leqslant p(n+1)-1.
\end{align*}
When $k\geq rn$, we obtain:
\begin{equation*}
    \frac{p(k+1)}{(1-p)(n-k)}\leqslant 1\Leftrightarrow k\leqslant n-p(n+1).
\end{equation*}
Therefore,
\begin{enumerate}
\item For states below $r$, the peak is attained at
\begin{equation*}
      \hat{c}_1 = \frac{\hat{k}_1}{n}, \text{ with } \hat{k}_1 = \lceil p(n+1)\rceil -1.
\end{equation*}
Observe that when $n$ is large, this yields $\hat{c}_1\approx p$.
\item For states above $r$, the peak is attained at
  \[
\hat{c}_2 = \frac{\hat{k}_2}{n}, \text{ with } \hat{k}_2 = n - \lfloor p(n+1)\rfloor.
\]
When $n$ is large, we obtain $\hat{c}_2 \approx 1-p$.
\end{enumerate}
Depending on the relative positions of $p$ and $r$, there can be one or two
peaks, as summarized as follows:
\begin{itemize}
    \item If $r\leqslant p$, $r\leqslant 1-p$: peak at $\hat{c}_{2}$,
    \item if $p\leqslant r\leqslant 1-p$: two peaks at $\hat{c}_1,\hat{c}_2$,
    \item if $1-p\leqslant r\leqslant p$: peak at $\frac{\lceil rn\rceil}{n}$, 
    \item $p\leqslant r$, $1-p\leqslant r$: peak at $\hat{c}_1$.
\end{itemize}
In the case there are two peaks, i.e., $p\leq r\leq 1-p$, let us find the relative heights of the peaks.
From (\ref{eq:pigeneral}), we find, assuming $rn\not\in\NN$, 
\begin{align*}
\pi(\lfloor rn\rfloor) &=
\pi\left(\hat{k}_1\right)\left(\frac{p}{1-p}\right)^{\lfloor rn\rfloor-\lceil p(n+1)\rceil+1}\\
&\times \frac{(n-\lfloor
  rn\rfloor+1)\cdots(n-\lceil p(n+1)\rceil +1)}{\lfloor rn\rfloor\cdots \lceil p(n+1)\rceil} 
\\[2ex]
\pi(\lfloor rn\rfloor+1) &= \pi\left(\hat{k}_2\right)\left(\frac{p}{1-p}\right)^{n-\lfloor
  p(n+1)\rfloor-\lfloor rn\rfloor-1}\\
&\times\frac{(\lfloor rn\rfloor+2)\cdots(n-\lfloor
    p(n+1)\rfloor)}{(n-\lfloor rn\rfloor-1)\cdots(\lfloor p(n+1)\rfloor+1)}
\\[2ex]
\pi(\lfloor rn\rfloor+1) &= \pi(\lfloor rn\rfloor)\frac{n-\lfloor rn\rfloor}{\lfloor rn\rfloor +1}
\end{align*}
Hence, assuming $p(n+1)\not\in\NN$,
\begin{equation}
    \frac{\pi\left(\hat{k}_2\right)}{\pi\left(\hat{k}_1\right)} = \left(\frac{p}{1-p}\right)^{2\lfloor rn\rfloor-n+1}.
    \label{eq:ratio}
\end{equation}
When $n$ is large, we obtain
\begin{equation}
    \frac{\pi\left(\hat{k}_2\right)}{\pi\left(\hat{k}_1\right)}\approx \left(\frac{p}{1-p}\right)^{n(2r-1)+1}.
\end{equation}

Observe that the peaks have equal heights when $p=0.5$, and when $r=0.5$, the
ratio is equal to $p/(1-p)$. 

\begin{figure}[htp]
    \includegraphics[width=\linewidth]{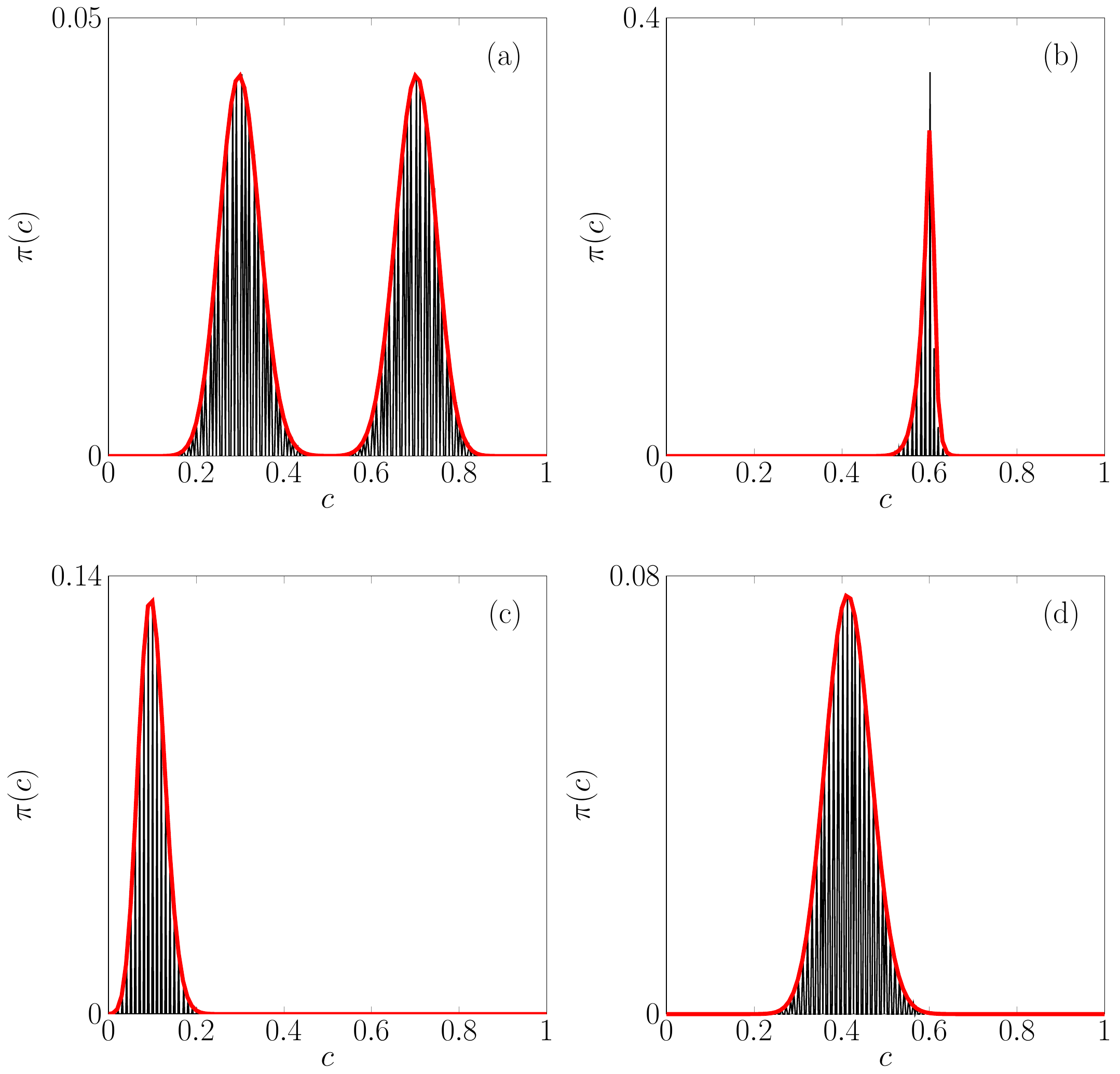}
    \caption{Stationary distributions of visited states for model with one threshold (upper row) with parameters (a) $r = 0.5$; $ p = 0.3$, (b) $r = 0.6$; $p=0.7$ and model with beta distribution (bottom row) with parameters (c) $\alpha = 8$; $\beta = 5$; $p = 0.1$, (d) $\alpha = 3$; $\beta = 1$; $p = 0.4$. Solid red lines are distributions obtained with Markov approach, black histograms are obtained with trajectories from Monte Carlo simulations for system of size $n = 100$ and thermalization time $t = 2\cdot 10^{6} $ MCS from 100 initial conditions evenly distributed on $[0,1]$ interval, averaged over 1000 independent runs.}
    \label{fig:hists}
\end{figure}

Besides, we have solved numerically by SCILAB the system of equations (\ref{eq:Pmatrix}), which is possible for reasonable values of $n$, and obtained  its solution $\pi(k)$, $k=0,\ldots,n$. 
Table~\ref{table:2} shows the value of the ratio of the two peaks for various values of $p,r$ as given by Eq. (\ref{eq:ratio}), compared to the output of SCILAB. 
Fig.~\ref{fig:hists} shows the computed distribution $\pi$ for $n=100$ for the one threshold case and also the case of the Beta distribution, compared to the histograms obtained from Monte Carlo simulations. 

\begin{table}[h!]
\centering
\begin{tabular}{|c|c||c|c|c|c|} 
 \hline
 $p$ & $r$ & $\pi\left(\hat{k}_2\right)/\pi\left(\hat{k}_1\right)$ &  $\pi\left(\hat{k}_1\right)$ &  $\pi\left(\hat{k}_2\right)$ & ratio \\ [0.5ex] 
 \hline\hline
  $0.21$ & $0.41$ & $3.7619048$ & $0.064177$ & $0.241429$ & $3.7619052$ \\
  $0.21$ & $0.61$ & $0.0187835$ & $0.2966234$ & $0.005572$ & $0.0187834$\\
  $0.25$ & $0.5$ & $0.3333333$ & $0.218683$ & $0.0728942$ & $0.3333334$\\ 
  \hline
\end{tabular}
\caption{Example of results for $n = 10$ for different values of $p$ and $r$ under the condition $p\leqslant r\leqslant 1-p$. In the table are presented the theoretical ratio given by Eq. (\ref{eq:ratio}) (left column), as well as the values  $\pi\left(\hat{k}_1\right)$ and $\pi\left(\hat{k}_2\right)$ computed numerically and the ratio between them (three rightmost columns).}
\label{table:2}
\end{table}

We comment on these results.
 The Markov approach permits to obtain the stationary probability distribution of the different states, for any value of $n$, without approximation. It is found that in the long run, even if any state has a nonzero probability to be reached, some states have a much higher probability than the others to appear. In the case of one threshold, we have analytically proved the presence of one or two peaks, and their positions when $n$ is large perfectly coincides with what was predicted by the mean field approach. It is complementary to the results given by the mean-field approach, since the Markov approach is able to give the probability of occurrence of each stationary state. On the other hand, the complexity of the system of linear equations (\ref{eq:Pmatrix}) induced by the Markov chain makes this approach not always tractable (e.g., with the Beta distribution). Nevertheless, we have shown that for reasonably large values of $n$ (e.g., $n=100$), this linear system can be solved numerically, giving a perfect fit with theoretical values, as shown by Table~\ref{table:2} and with Monte Carlo simulations as well, see Fig~\ref{fig:hists}.

\section{Summary and research directions for the future}
In this paper, we investigated the threshold model with anticonformity under asynchronous update mode, which mimics continuous time. We considered two cases: (1) homogeneous, in which all agents had the same threshold and (2) heterogeneous, in which the thresholds are given by the beta distribution function. The homogeneous case with $r=0.5$ is identical to the homogeneous symmetrical threshold model with anticonformity \cite{Now:Szn:19}. Moreover, it is almost identical to the majority-vote process \cite{Lig:85,Oli:92}. The only difference between the models is when the number of active and inactive agents in the neighborhood of a chosen agent is equal. In such a case, the state of the system does not change within the majority-vote model, whereas within the threshold model the change is possible. From this point of view, the threshold model with anticonformity under asynchronous updating can be treated as a generalization of a majority-vote model. 

On the complete graph, the homogeneous threshold model does not give particularly interesting results. The relationship between the stationary ratio of active agents and the probability of anticonformity consists of linear dependencies, similarly as for the homogeneous symmetrical threshold model \cite{Now:Szn:19,Now:Szn:21}. The only interesting feature of this model is the discontinuity that appears at $c=r=1-p$. In the result, the system reaches one of two different steady states, depending on the initial conditions. Much richer behavior is observed in the heterogeneous model with thresholds given by the beta distribution function, parametrized by $\alpha, \beta$, which allows tuning the model to the homogeneous one ($\alpha,\beta \rightarrow \infty$) to maximally heterogeneous (i.e. described by the uniform distribution function). 

A particularly interesting behavior is obtained if at least one of the shape parameters $\alpha$ or $\beta$ is larger than one and both parameters are finite. In this case PDF has a shape that reminds those of actual trait manifestation in behavior, as reported by psychologists \cite{Fle:Gal:09}, i.e., uni modal, not necessarily symmetrical, function with maximum at the value $0<r<1$. In such a case a phase transition appears, which is continuous for $\alpha = \beta$, and discontinuous otherwise. In the latter case, the transition involves phenomena typical of social systems, such as social hysteresis \cite{Sche:Wes:Bro:03} and the critical mass \cite{Cen:etal:18}.

The future research on the model can be conducted in several directions, related to the following questions:
\begin{itemize}
    \item How the results would change if the threshold for anticonformity would be different than that for conformity? This question is inspired by the work on the  $q$-voter model with generalized anticonformity \cite{Abr:etal:21}. In the $q$-voter model such a generalization resulted in switching from continuous to discontinuous phase transitions for some values of parameters. The question is if the same phenomena would be observed for the threshold model.
    \item How the structure of a network would influence the results? This question is inspired by the work on the symmetrical threshold \cite{Now:Szn:21}. It was shown that on random graphs with the degree observed empirically for social networks, the  largest social hysteresis is observed for $r \in (0.65,0.85)$. This was a meaningful result from the social point of view and thus it would be desirable to check if it appears also in the asymmetric model studied here.
    \item How the results would change if the quenched approach to anticonformity would be used. In this version of the model, we used the annealed approach, in the sense that each agent could anticonform (with probability $p$) or conform (with probability $1-p$). However, we could use also the quenched approach, in which a fraction $p$ of agents are permanently anticonformists. This question is inspired by the work on the  $q$-voter model with nonconformity under quenched and annealed approaches \cite{Jed:Szn:17}. It was shown that on the complete graph both approaches give the same result for the $q$-voter model with anticonformity, whereas different for the model with independence. The question is to what extend this result is universal.
\end{itemize}

\begin{acknowledgments}
This research was supported by the National Science Center (NCN, Poland) grant number 2019/35/B/HS6/02530.
\end{acknowledgments}


%

\end{document}